\documentclass[prl,twocolumn,showpacs,preprintnumbers,amsmath,amssymb,superscriptaddress]{revtex4}
\pdfoutput=1
\usepackage{graphicx}
\usepackage{dcolumn}% Align table columns on decimal point
\usepackage{bm}% bold math

\usepackage{epstopdf}

\usepackage{color}

\begin{document}

\title{Magnetic polarization of the americium $J = 0$ ground state in AmFe$_{2}$}

\author{N. Magnani}
\affiliation{European Commission, Joint Research Centre (JRC), Institute
for Transuranium Elements (ITU), Postfach 2340, D-76125 Karlsruhe, Germany}

\author{R. Caciuffo}
\affiliation{European Commission, Joint Research Centre (JRC), Institute
for Transuranium Elements (ITU), Postfach 2340, D-76125 Karlsruhe, Germany}
\email{roberto.caciuffo@ec.europa.eu}

\author{F. Wilhelm}
\affiliation{European Synchrotron Radiation Facility (ESRF), B.P.220, F-38043 Grenoble, France}

\author{E. Colineau}
\affiliation{European Commission, Joint Research Centre (JRC), Institute
for Transuranium Elements (ITU), Postfach 2340, D-76125 Karlsruhe, Germany}

\author{R. Eloirdi}
\affiliation{European Commission, Joint Research Centre (JRC), Institute
for Transuranium Elements (ITU), Postfach 2340, D-76125 Karlsruhe, Germany}

\author{J.-C. Griveau}
\affiliation{European Commission, Joint Research Centre (JRC), Institute
for Transuranium Elements (ITU), Postfach 2340, D-76125 Karlsruhe, Germany}

\author{J. Rusz}
\affiliation{Department of Physics and Astronomy, Uppsala University, Box 516, S-75120 Uppsala, Sweden}

\author{P. M. Oppeneer}
\affiliation{Department of Physics and Astronomy, Uppsala University, Box 516, S-75120 Uppsala, Sweden}

\author{A. Rogalev}
\affiliation{European Synchrotron Radiation Facility (ESRF), B.P.220, F-38043 Grenoble, France}

\author{G. H. Lander}
\affiliation{European Commission, Joint Research Centre (JRC), Institute
for Transuranium Elements (ITU), Postfach 2340, D-76125 Karlsruhe, Germany}

\date{\today}

\begin{abstract}
Trivalent americium has a non-magnetic ($J$ = 0) ground state arising from the cancelation of the orbital and spin moments. However, magnetism can be induced by a large molecular field if Am$^{3+}$ is embedded in a ferromagnetic matrix. Using the technique of x-ray magnetic circular dichroism, we show that this is the case in AmFe$_2$. Since $\langle J_z \rangle$ = 0, the spin component is exactly twice as large as the orbital one, the total Am moment is opposite to that of Fe, and the magnetic dipole operator $\langle T_{z} \rangle$ can be determined directly; we discuss the progression of the latter across the actinide series.
\end{abstract}


\maketitle

%\section{Introduction}

Ordered magnetism is a result of spin-polarization of electrons, but there are two elements in the periodic table in which the intrinsic magnetic moment is zero despite the electrons being spin-polarized: europium and americium. For the free Eu$^{3+}$ and Am$^{3+}$ ions the $f$-electron count $n_f$ is six, the spin and orbital moments have the same magnitude and opposite direction, and the resulting $J = 0$ ground state is non-magnetic. Nevertheless, since a large spin polarization is present, the application of a magnetic field can induce a moment by $J$-mixing with excited states. A textbook example of this phenomenon is the temperature-independent paramagnetism observed in several Eu$^{3+}$ compounds \cite{vanvleck68}. Europium, however, has a tendency towards the magnetic $f^{7}$ (divalent) configuration \cite{ruck11} and loses its magnetism only under high pressure \cite{debessai09}. Americium, in contrast, exhibits a stable trivalent oxidation state and, like most of its compounds, shows temperature independent susceptibility and no ordered magnetism \cite{edelstein06}.

One can expect that long-range order of the moments induced in the virtually nonmagnetic sublattice \cite{johannes05} will be evident when these ions are embedded in a strong ferromagnetic matrix, because of the large molecular field created by the exchange interaction \cite{wolf60}. This will be the case, for instance, in the cubic Laves phase AmFe$_2$, a compound where the $f$-$d$ exchange interaction is anticipated to be very large \cite{liu91} and ferromagnetic order is observed already at room temperature \cite{lander77}. We have therefore used x-ray magnetic circular dichroism (XMCD), which is an element- and shell-specific technique, to study the size and nature of the spin and orbital moments induced by the exchange field on Am$^{3+}$ in AmFe$_2$. In order to determine the spin component of the magnetic moment ($\mu_S = -2 \langle S_z \rangle$) from XMCD measurements it is necessary to know the expectation value of the magnetic dipole operator $\bf{T} = \sum_i[\bf{s}_i-3\bf{r}_i(\bf{r}_i \cdot \bf{s}_i)/r_i^2]$ (which is associated with the spin-dependent asphericity of the electronic cloud \cite{collins95}) because the sum rules only give the value of $\langle S_{\rm{eff}} \rangle \equiv \langle S_z \rangle + 3 \langle T_z \rangle$ \cite{thole92,carra93}. In some favorable cases $\langle T_z \rangle$ can be assessed by using a complementary method to estimate the total magnetic moment of the absorbing atom, but in general it is not easily accessible nor understood \cite{moore09}; for instance, early studies on UFe$_2$ \cite{finazzi97} assumed that $\langle T_z \rangle = 0$ (as is standard practice for itinerant $3d$ ferromagnets), whereas in NpOs$_2$ and PuFe$_2$ its value is consistent with the one calculated in the intermediate coupling (IC) scheme \cite{wilhelm13}.
In the case of AmFe$_2$, we are in the unique position of determining $\langle T_z \rangle$ directly from the dichroic signals measured at the M$_4$ and M$_5$ Am absorption edges because $\langle J_z \rangle$ remains zero (and therefore $\langle L_z \rangle = - \langle S_z \rangle$)
\emph{even in the magnetically-induced state}.

%\section{Experimental results}

\begin{figure}
\includegraphics[width=7.0cm]{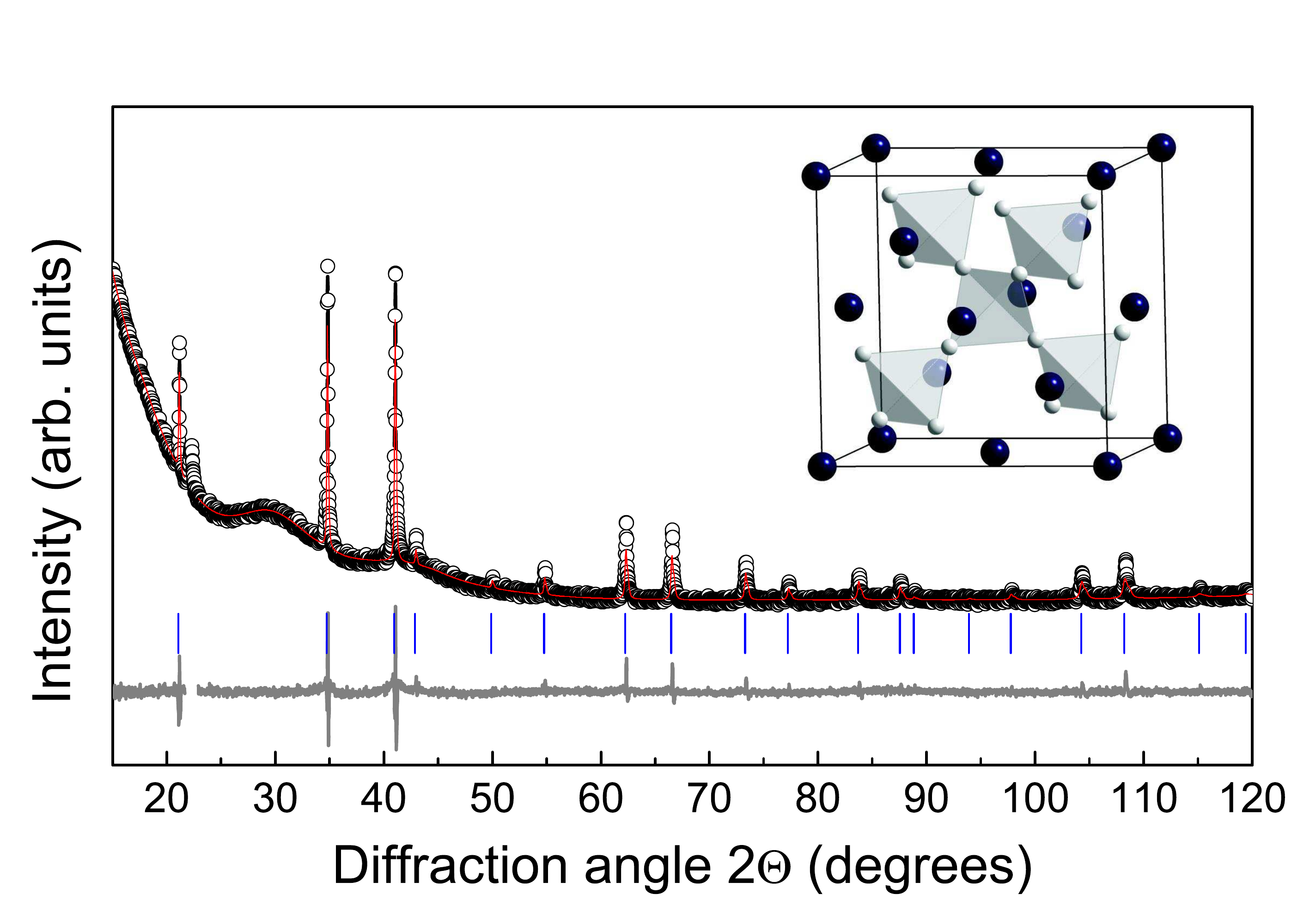}
\caption{Observed (circles), calculated (full red line), and difference (lower trace) x-ray diffraction pattern recorded at room temperature for the AmFe$_2$ sample used in this study. Vertical ticks indicate the position of Bragg peaks. The broad peaks at low angles are due to the sample holder. Inset: Cubic C-15 structure of the Laves phase AmFe$_{2}$ (Space Group Fd$\overline{3}$m, room temperature lattice constant $a_0$ = 7.300 \AA). Am atoms are represented by dark large spheres, Fe atoms by smaller spheres.
\label{structure}}
\end{figure}

\begin{figure}
\includegraphics[width=6.5cm]{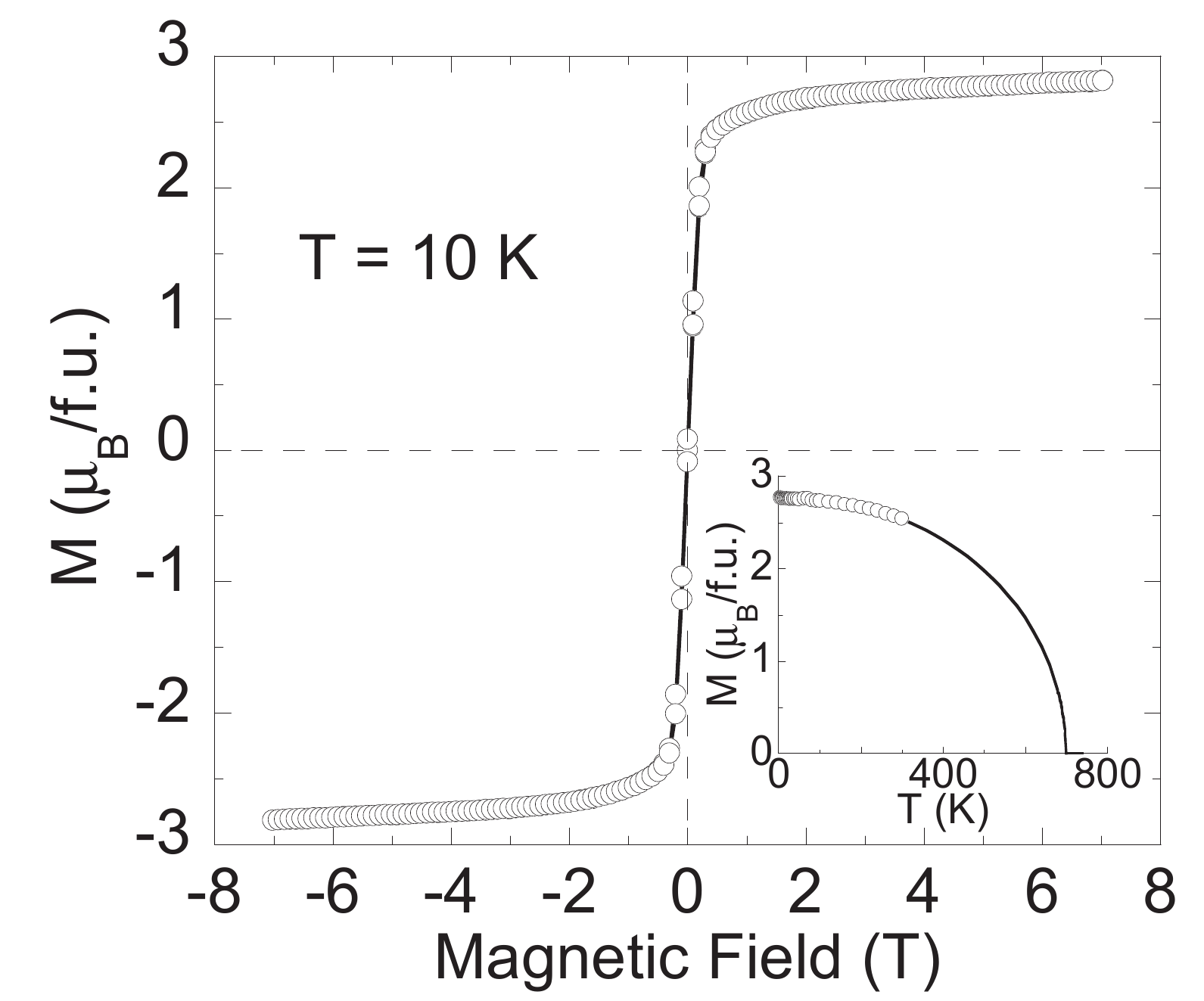}
\caption{Magnetization curve measured for AmFe$_{2}$ at 10 K. The inset shows the temperature dependence of the magnetization M (open circles) fitted by a $J = 5/2$ Brillouin function (solid line), providing an estimate of the Curie temperature $T_{C} \sim 700~\rm{K}$.
 \label{magnetization}}
\end{figure}

AmFe$_2$ was fabricated  by arc melting stoichiometric amounts of elemental constituents on a water-cooled copper hearth, under Ar (6N) atmosphere. A Zr alloy was used as an oxygen getter. The weight losses were examined after arc melting and resulted to be less than 0.5~\%. The sample was melted 5 times and crushed before the last melt, to ensure complete homogeneity of the alloy button. X-ray diffraction analysis performed at room temperature confirmed that the sample obtained was single-phase with the C-15 cubic structure (Fig.~\ref{structure}), with a lattice parameter in agreement with earlier work \cite{lander77}. Magnetization experiments with a Quantum Design MPMS-7 SQUID magnetometer were carried out between 2 and 300 K and showed the sample to be ferromagnetic in the whole temperature range, again as expected. The observed magnetization curve is typical for a soft ferromagnet with vanishing coercive field. At $T = 10~\rm{K}$, the saturation moment per formula unit is $2.8(1) \mu_B$, as shown in Fig.~\ref{magnetization}. Its temperature dependence in a field of 7 tesla was fitted to a $J = 5/2$ Brillouin function (Fig.~\ref{magnetization}, inset) that provides an estimate of the Curie temperature $T_C \sim 700~\rm{K}$, in good agreement with previous estimates \cite{lander77}.

\begin{figure}
\includegraphics[width=8.0cm]{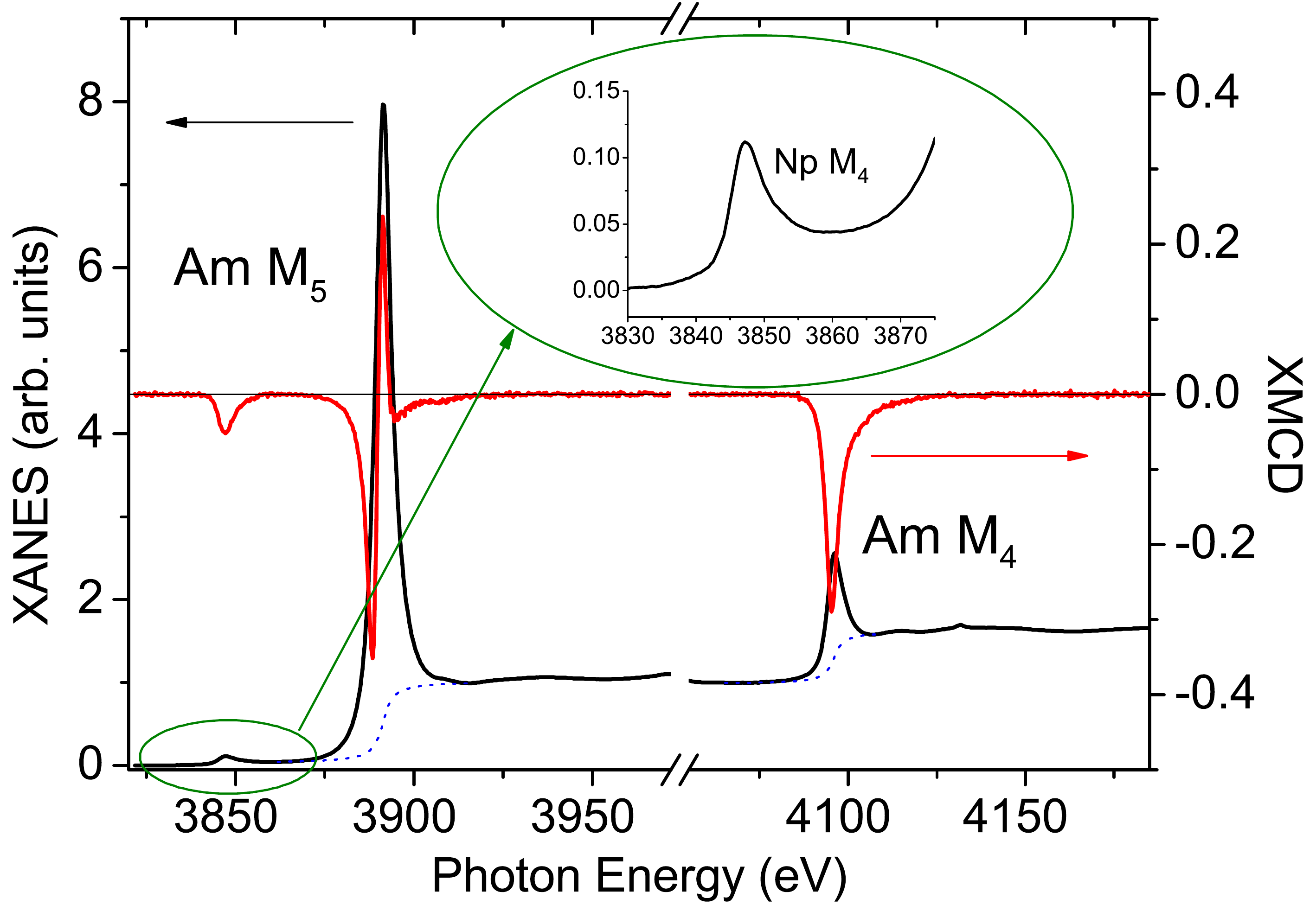}
\caption{(Color online) The x-ray absorption (XANES) and x-ray magnetic circular dichroism (XMCD) spectra as a function of photon energy through the Am M$_5$ and M$_4$ edges in AmFe$_2$. The experiment was conducted at room temperature in an applied field of 3 tesla. The spectra have been corrected for self-absorption effects and incomplete circular polarization of the incident beam.
The inset shows the XANES signal from the Np M$_4$ edge.  $^{237}$Np is a decay product of $^{241}$Am and the gamma spectra from the sample showed this to be present at the $\sim$1\% level (Np in NpFe$_2$ is strongly magnetic --- see \cite{wilhelm13}).
 \label{xmcd}}
\end{figure}

The x-ray-absorption-spectroscopy (XAS) and XMCD experiments were carried out at the European Synchrotron Radiation Facility (ESRF) using the ID12 beamline on a 16 $\mu$g sample of AmFe$_2$ ($\sim40~\times 200~\times 150~\mu \rm{m}$) taken from the batch made at ITU and encapsulated in an Al holder with kapton windows of 60 $\mu$m thickness in total. Data were collected at room temperature across the M$_{4,5}$ edges of Am in the photon energy range 3.830-4.170 keV. Saturation was already obtained for a field of 0.5 T, consistent with magnetization measurements.  The spectra recorded using the total-fluorescence-yield detection mode in backscattering geometry with a 3 T magnetic field are shown in Fig.~\ref{xmcd}. The integrated white line XAS intensities $I_{M_{4,5}}$ measured for the two photon helicities and the XMCD spectra were obtained as discussed in the Supplementary Material \cite{SM,S1,S2,S3,S4,S5,S6,S7}.

\begin{figure}
\includegraphics[width=6.0cm]{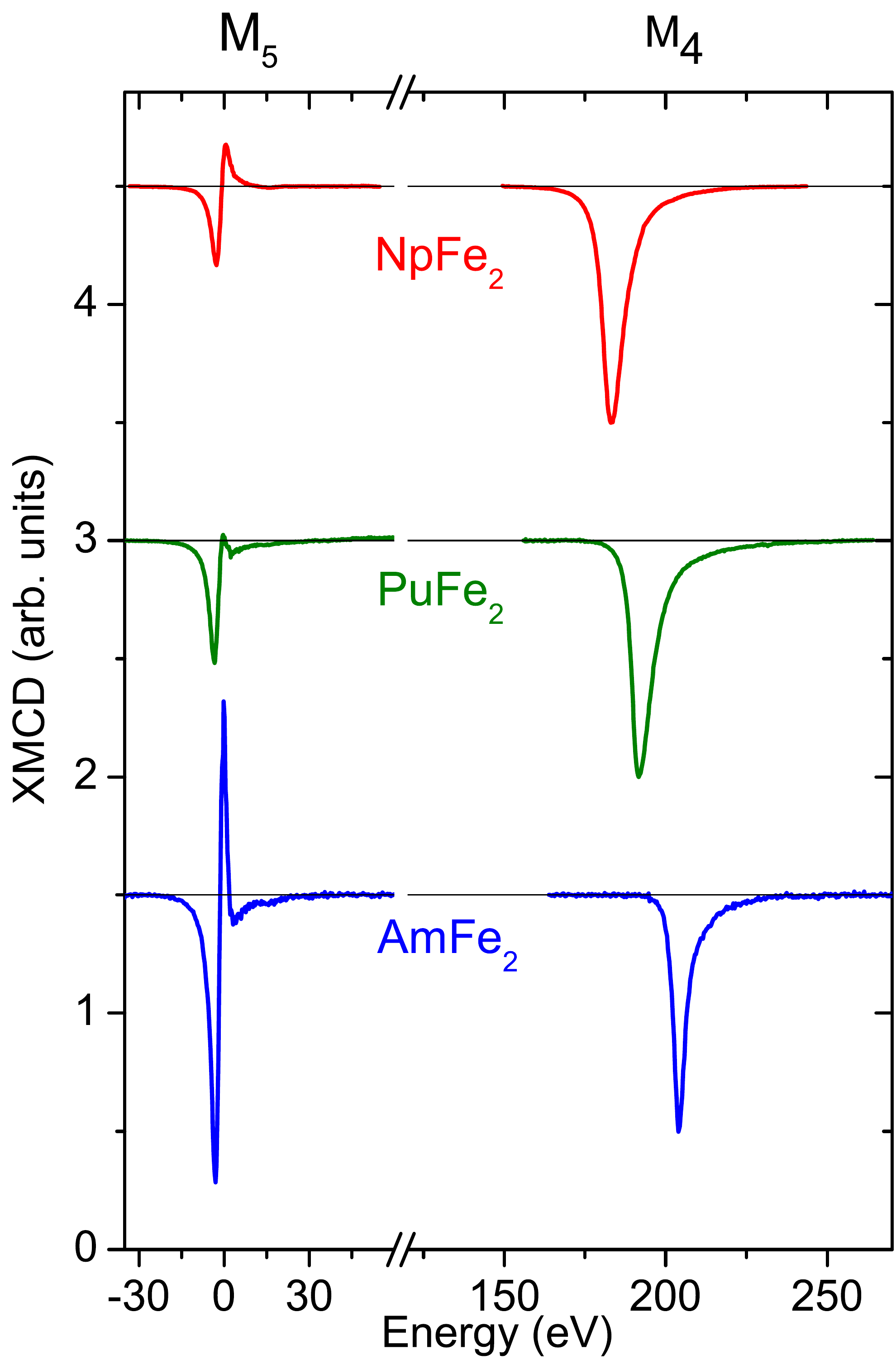}
\caption{(Color online) XMCD spectroscopic shapes for the An M$_5$ and M$_4$ edges of AnFe$_2$ compounds (Np through Am; NpFe$_2$ and PuFe$_2$ from \cite{wilhelm13}, AmFe$_2$ --- this work) with the energy of the M$_5$ edge taken as zero and the amplitude of the M$_4$ edge normalized to unity. Note that although for any one element the two signals are correctly represented, there is no scaling between the signals for different elements. For example, the signal for AmFe$_2$ is much smaller in absolute terms than that found for the large Np moment in NpFe$_2$. The narrow linewidth of the M$_4$ XMCD signal for Am in AmFe$_2$ (about 50\% of that found for the actinides in NpFe$_2$ and PuFe$_2$) is consistent with the assumption of localized $5f$ states, and found also in PuSb \cite{janoschek15}.
 \label{xmcdseries}}
\end{figure}

The so-called XAS branching ratio $B = I_{M_{5}}/(I_{M_{5}}+I_{M_{4}})$ is proportional to the expectation value of the angular part of the valence states spin-orbit operator $\langle {\bf l} \cdot {\bf s} \rangle = (3/2) n_{7/2} - 2 n_{5/2}$ \cite{thole88},
\begin{equation}\label{so}
\frac{2\langle {\bf l} \cdot {\bf s} \rangle}{3(14-n_{f})}-\Delta = -\frac{5}{2}(B-\frac{3}{5})
\end{equation}
where $\Delta$ is a quantity dependent on the electronic configuration ($\Delta \sim$ 0.005 for Am$^{3+}$) \cite{vanderlaan04}.
We find $B = 0.88(4)$, which is close to the value ($B = 0.93$) expected for a $5f^6$ configuration assuming IC \cite{vanderlaan96} confirming that there are six $5f$ electrons with the majority ($n_{5/2} = 4.95$) in the $j = 5/2$ subshell. The XMCD spectra are composed of a small down-up feature at the M$_5$ edge and a large (negative) peak at M$_4$. This is the typical spectral shape characteristic of a dominating orbital moment with an oppositely oriented spin, as observed for lighter actinide (An) compounds \cite{halevy12,hen14} and in particular for the Np- and Pu-based AnFe$_2$ Laves phases \cite{wilhelm13} (see Fig. \ref{xmcdseries}). A completely different shape is expected for the $n_f = 7$ configuration, for which the small orbital moment implies a positive dichroic signal at the M$_4$ edge, opposite in sign to a M$_5$ peak of almost equal intensity, as seen for Eu$^{2+}$ \cite{ruck11} and Cm$^{3+}$ \cite{wilhelm15}.

%\section{Discussion}

The orbital contributions to the magnetic moment carried by the Am atoms can be determined by the sum rule \cite{thole92}
\begin{equation}\label{om}
\langle L_{z}\rangle = \frac{14-n_{f}}{I_{M_{5}}+I_{M_{4}}} (\Delta I_{M_{5}}+\Delta I_{M_{4}})
\end{equation}
where $\Delta I_{M_{5}}+\Delta I_{M_{4}}$ is the total dichroic signal at the Am $M_{4,5}$ edges. Applying this sum rule, we obtain the orbital moment on Am as $\mu_L=-\langle L_{z}\rangle = + 0.44(5) \mu_B$ (the positive sign indicates that the orientation is parallel to the moment of the Fe sublattice, as found in all other An Laves phases \cite{wilhelm13}). A second sum rule correlates the measured dichroic signal and the spin polarization $\langle S_{z} \rangle$, stating that \cite{carra93}
\begin{equation}\label{sm}
\langle S_{\rm{eff}} \rangle \equiv \langle S_{z}\rangle + 3\langle T_{z}\rangle = \frac{14-n_{f}}{2(I_{M_{5}}+I_{M_{4}})} (\Delta I_{M_{5}}-\frac{3}{2}\Delta I_{M_{4}})
\end{equation}
The experimental data for AmFe$_2$ provide $\langle S_{\rm{eff}} \rangle = -0.135(15)$.

The key point to understand our experimental data is that the ground state of the Am$^{3+}$ ion maintains its $\langle J_{z} \rangle$ = 0 character even though $\langle J^{2} \rangle$ becomes different from zero. This is because in the present case J-mixing is almost entirely due to $f$-$d$ exchange (which does not commute with \textbf{J}$^{2}$ but does commute with $J_z$), whereas the role of the crystal field is negligible. To prove this statement, we summarize below the spectroscopic properties of Am$^{3+}$ ions in AmFe$_2$. The single-ion Hamiltonian which describes its 5$f$ electronic states can be written as ${\cal H} = {\cal H}_{\rm{FI}} + {\cal H}_{\rm{CF}} + {\cal H}_{\rm{Z}} + {\cal H}_{\rm{ex}}$, where the four main contributions are the free-ion Hamiltonian ${\cal H}_{\rm{FI}} = \sum_{k=1}^3 F^{2k} f_{2k} + \zeta_{5f} \sum_{i=1}^{n_f} \bf{l}_i \cdot \bf{s}_i$, the crystal-field Hamiltonian ${\cal H}_{\rm{CF}} = B_4^0 [C_0^{(4)} + \sqrt{5/14}(C_{-4}^{(4)}+C_{4}^{(4)})] + B_6^0 [C_0^{(6)} - \sqrt{7/2} (C_{-4}^{(6)}+C_4^{(6)})]$, the Zeeman term ${\cal H}_{\rm{Z}} = -\mu_B \bf{H} \cdot (\bf{L} + 2 \bf{S})$, and the $f$-$d$ exchange interaction, represented by an internal field $\bf{H}_{\rm{int}}$ which is generated by the ordered Fe sublattice and acts only on the Am spin (${\cal H}_{\rm{ex}} = -2 \mu_B \bf{H}_{\rm{int}} \cdot \bf{S}$). All the operators and symbols used are defined in \cite{wybourne65}. Following Hund's rules, the $^7 F$ spectroscopic term minimizes the Coulomb repulsion energy, and the spin-orbit interaction selects the $J=0$ singlet as the free-ion ground state, with the first excited manifold having $J = 1$ and lying between $\sim$ 220 and 340 meV \cite{carnall92,apostolidis10}. The cubic crystal field potential has no effect on these two multiplets, as it splits only those with $J \geq 2$. Moreover, the non-axial part of ${\cal H}_{\rm{CF}}$ can only mix the $J_z = 0$ ground state with $J_z = \pm 4$  components of excited multiplets; as the lowest $J = 4$ manifold has an energy of about 1.2 eV, we can safely neglect this contribution. The quantization axis $z$ is therefore selected by the direction of the internal field and $\langle J_z \rangle \simeq 0$, provided that $H_{\rm{int}} \gg H$. This is a reasonable assumption since $H_{\rm{int}} \simeq 180~\rm{T}$ was estimated by rescaling the value proposed for the isostructural lanthanide (Ln) series LnFe$_2$ \cite{radwanski86} to account for the different expectation value of $\langle r^{2} \rangle$ between the radial $5f$ and $4f$ wavefunctions. The only other relevant parameters present in ${\cal H}$ are the Slater integrals $F^{2k}$ (that we fixed to the values given for IC in \cite{moore09}) and the spin-orbit parameter $\zeta_{5f}$ (that we adjusted to 285 meV in order to reproduce the experimental value of the branching ratio $B$). A full diagonalization of ${\cal H}$ provides the expectation value $\langle L_z \rangle = -0.47$ for the ground state, in excellent agreement with the experimental determination.

The fact that $\langle J_z \rangle$ = 0, a result that as shown above is independent from the computational details, allows us to determine $\langle S_z \rangle = - \langle L_z \rangle$ directly from the first sum rule (Eq. \ref{om}). We obtain $\mu_S = - 2 \langle S_{z} \rangle =  -0.88(10)~\mu_B$, and therefore $\mu_{\rm{Am}}=\mu_L+\mu_S =  -0.44(11)~\mu_B$ in remarkable agreement with neutron diffraction experiments \cite{lander77}. From the bulk saturation moment (see Fig. \ref{magnetization}) we obtain for the Fe sites a moment $\mu_{\rm{Fe}} =  1.6(1)~\mu_B$, within experimental errors equal to the one observed for analog LnFe$_2$ series but almost three times larger than in UFe$_2$ \cite{lebech89}, which is a well known itinerant system with strong $f$-$d$ hybridization.

Determining the above values required no assumptions on $\langle T_z \rangle$, which can then be obtained independently knowing the value of $\langle S_z \rangle$ and the experimental value of the ratio $\langle L_{z} \rangle$/$\langle S_{\rm{eff}} \rangle$ derived from the sum rules (Eqs. \ref{om}-\ref{sm}). This gives $3\langle T_z \rangle = -0.57(5)$ and a ratio $r = 3 \langle T_z \rangle / \langle S_z \rangle = -1.3(2)$. With the model described above, we calculate $3 \langle T_z \rangle = -0.51$ and $\langle S_z \rangle = 0.47$, corresponding to $r = -1.1$.

\begin{figure}
\includegraphics[width=7.0cm]{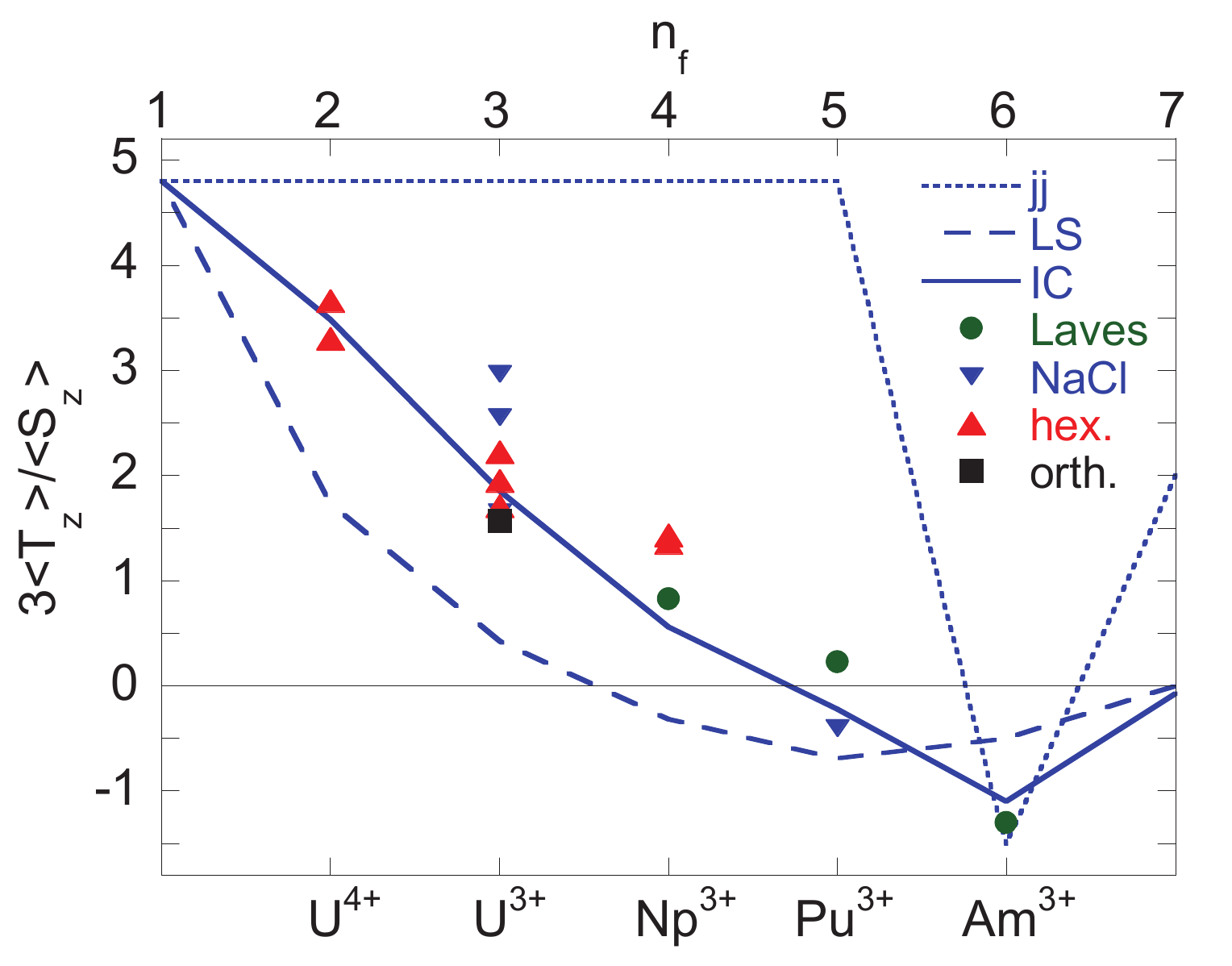}
\caption{(Color online) Ratio (r) between the expectation values of the magnetic dipole operator $3 \langle T_z \rangle$ and the spin operator $\langle S_z \rangle$ as a function of the 5f shell occupation number $n_{f}$. Experimental data for different compounds are represented by symbols identifying their crystallographic structure (circles for C-15 cubic Laves phases, squares for NaCl-type structure, and triangles for hexagonal lattice groups). Calculated values are shown for $LS$ (dashed line), $jj$ (dotted line), and IC (solid line). All the calculated values refer to the free-ion ground state, except for the $f^6$ configuration, where $J$-mixing with the first excited state is taken into account; otherwise $r$ cannot be determined since $\langle T_z \rangle = \langle S_z \rangle = 0$.
 \label{ratiomS}}
\end{figure}

The values for the $r$ ratio for different An compounds (UGe$_2$ \cite{kernavanois01b,okane06b,sakurai06}, US \cite{kernavanois01}, USe, UTe \cite{okane06}, USb$_{0.5}$Te$_{0.5}$ \cite{dalmas97}, UPd$_2$Al$_3$ \cite{yaouanc98}, UNi$_2$Al$_3$ \cite{kernavanois00}, UCoAl, UPtAl \cite{kucera02}, URhAl \cite{grange98}, NpNi$_5$ \cite{hen14}, Np$_2$Co$_{17}$ \cite{halevy12}, NpOs$_2$, PuFe$_2$ \cite{wilhelm13} and PuSb \cite{janoschek15}) have been derived from data reported in the literature and are shown in Fig. \ref{ratiomS}, where they are compared with theoretical values calculated in IC, as well as in the two limit coupling schemes $LS$ and $jj$. The value we obtained for AmFe$_2$ is very close to the IC curve and follows the experimental trend observed for the other Laves phases. We stress that all the symbols presented in Fig. \ref{ratiomS} refer to cases where a \emph{purely experimental} determination of $r$ could be performed by combining the values of $\langle L_z \rangle$ and $\langle S_{\rm{eff}} \rangle$ deduced from XMCD spectra and the total magnetic moment \emph{on the An site} measured independently by neutron diffraction or M\"{o}ssbauer spectroscopy, without any input from electronic structure calculations.

%The only notable exception to this behavior might be UFe$_2$ for which a vanishing $\langle T_z \rangle$ value has been assumed in %treating the XMCD results \cite{finazzi97}. However, the results for the M$_{4,5}$ edges given in \cite{finazzi97} are at odds with those %later reported for the N$_{4,5}$ edges of the same compound \cite{okane06}, so that no conclusive assignment can be made.

To complement the single-site model treatment of the Am magnetism, we have performed density functional theory (DFT) calculations of AmFe$_2$ using the DFT+$U$ method.
%Ref.~\cite{wilhelm13} contains a detailed description of our computational approach.
%
%Strong electronic correlation effects often pose a challenge to DFT calculations, and this turned out to be the case for the AmFe$_2$ %system, too. We found that  DFT+$U$ calculations employing the around mean-field approach \cite{ldauamf} with low values of $U < %1$~eV
%provided XAS and XMCD spectra having  qualitatively the same shape as the measured ones, yet only for $U < 0.5$ eV  the branching ratio %approached the experimental one.
%Also, for $U \approx 1$~eV the calculated ground state magnetic spin and orbital moment are comparable to the measured ones.
The results are summarized in the Supplementary Material \cite{SM2,S10,S11,S12,S13,S14,S15,S16,S17,S18}.
An improved treatment  of AmFe$_2$, whose electronic structure involves an interplay of strong correlation effects with exchange fields from the ordered Fe sublattice, and strong spin-orbit interaction of Am, could require a computational scheme going beyond the present static DFT+$U$ approach.

%\section{Conclusions}
In conclusion, by an XMCD experiment on AmFe$_2$ we have directly observed the ordered magnetic moment induced on the formally $J=0$ ground state of Am$^{3+}$  by the exchange interaction with the ferromagnetically ordered iron sublattice; despite the absence of an intrinsic magnetic moment in the free ion, the fact that a large spin polarization is present results in a significant exchange interaction. Our result not only confirms the previous indication on the total Am moment from neutron diffraction experiments, but by probing the orbital and the effective spin moment separately, it allows us to attribute the resultant induced moment (\textit{antiparallel} to the Fe one) to significantly localized $5f$ electrons, a situation very different to the isostructural UFe$_2$.

The intrinsic relation $\langle S_{z} \rangle = -\langle L_{z} \rangle$ resulting from the uniaxial symmetry of the Hamiltonian offered us a unique opportunity to determine \emph{directly from the XMCD spectra} the expectation value of the magnetic dipole operator $\langle T_{z} \rangle$, an elusive quantity which is experimentally accessible only in a limited number of cases and normally requires a combination of several techniques. By comparing the value of $3 \langle T_{z} \rangle / \langle S_{z} \rangle$ for $n_f = 6$ to that of other $5f$ compounds with $n_f \leq 5$ we infer that this quantity is well described within a single-ion, intermediate coupling theory for all light actinides, in a way which is largely independent of their electronic (de)localization. This finding resolves a long-standing issue of what to use for $\langle T_{z} \rangle$ in interpreting XMCD experiments at the actinide M$_{4,5}$ (or N$_{4,5}$) edges. These studies address the orbital and spin moments that have been of interest since at least the 1980s \cite{brooks83}, and are still complicated today to determine theoretically \cite{pezzoli11}. Single-crystal samples, as usually needed for neutron experiments, are not required, and microgram scales are sufficient. This opens the way to future experiments to determine $\mu_L$ and $\mu_S$ in actinide systems from XMCD data only.

%\acknowledgments{
We thank D. Bou\"{e}xi\`{e}re, G. Pagliosa, and P. Amador Celdran, for their technical support in the preparation and encapsulation of the sample, and P. Colomp of the ESRF radioprotection services for his cooperation.
%}

\bibliography{amfexmcd}
\end{document}